\documentclass[useAMS,usenatbib]{mnras}
\usepackage{amssymb}
\usepackage{amsfonts}
\usepackage{tabularx}
\usepackage{graphicx}
\usepackage{ulem}
\usepackage{array}
\usepackage{color}
\usepackage{booktabs}
\usepackage[fleqn]{amsmath}
\usepackage{mathabx}
\makeatletter
\newcommand*{\rom}[1]{\expandafter\@slowromancap\romannumeral #1@}
\makeatother
\numberwithin{equation}{section}
\def\d{{\rm d}}
\def\e{{\rm e}}
\newcommand{\p}{\partial}

\newcommand{\case}{\textstyle\frac}

\newcommand{\ii}{{\mathrm i}}

\newcommand{\vk}{{\mathbf k}}

\newcommand{\il}{\int\limits}
\newcommand{\ilii}{\int\limits_{-\infty}^{\infty}}

\renewcommand{\[}{\left[}

\renewcommand{\Im}{\textrm{Im}\,}

\newcommand{\mfg}{{\mathfrak{g}}}
\newcommand{\mfB}{{\mathfrak{B}}}

\newcommand{\be}{\begin{equation}}
\newcommand{\ee}{\end{equation}}

\newcommand{\ba}{\begin{align}}
\newcommand{\ea}{\end{align}}

\begin{document}

\title[Damped perturbations in stellar systems]{Damped perturbations in stellar systems: \\Genuine modes and Landau-damped waves}

\author[E.\,V.\,Polyachenko et al.]{
E.~V.~Polyachenko,$^{1}$\thanks{E-mail: epolyach@inasan.ru}
I.~G.~Shukhman,$^{2}$\thanks{E-mail: shukhman@iszf.irk.ru}
O.~I.~Borodina,$^{1}$\thanks{E-mail: oborodina@inasan.ru}
\\
$^{1}$Institute of Astronomy, Russian Academy of Sciences, 48 Pyatnitskya St., Moscow 119017, Russia\\
$^{2}$Institute of Solar-Terrestrial Physics, Russian Academy of Sciences, Siberian Branch, P.O. Box 291, Irkutsk 664033, Russia
}


\maketitle
\pagerange{\pageref{firstpage}--\pageref{lastpage}} \pubyear{2020}

\maketitle

\begin{abstract}
This research was stimulated by the recent studies of damping solutions in dynamically stable spherical stellar systems. Using the simplest model of the homogeneous stellar medium, we discuss nontrivial features of stellar systems. Taking them into account will make it possible to correctly interpret the results obtained earlier and will help to set up decisive numerical experiments in the future. In particular, we compare the initial value problem versus the eigenvalue problem. It turns out that in the unstable regime, the Landau-damped waves can be represented as a superposition of van Kampen modes {\it plus} a discrete damped mode, usually ignored in the stability study. This mode is a solution complex conjugate to the unstable Jeans mode. In contrast, the Landau-damped waves are not genuine modes: in modes, eigenfunctions depend on time as $\exp (-\ii \omega t)$, while the waves do not have eigenfunctions on the real $v$-axis at all. However, `eigenfunctions' on the complex $v$-contours do exist. Deviations from the Landau damping are common and can be due to singularities or cut-off of the initial perturbation above some fixed value in the velocity space.
\end{abstract}


\section{Introduction}

It is well known that spherical systems, in contrast to stellar systems of other geometry, have a fair amount of stability \citep[e.g.,][]{1984sv...bookQ....F}. A fundamental result is the proof of the stability of isotropic spheres \citep{1960AZh....37..918A, 1962spss.book.....A, 1971PhRvL..26..725D}. In this vein, the choice of spheres for studying damped oscillations is obvious. However, the difficulties one encounters explain low activity on this topic.

The state-of-art was presented recently by \citet{2020MNRAS.492.6019H}. Utilizing correlation analysis, the authors attempt to reproduce the real part of frequency for a damped mode obtained earlier by \citet{1994ApJ...421..481W}. In the latter, using a matrix method, damped dipole and quadrupole modes for ergodic King models with parameters $W_0 = 3\,...\,7$ were obtained. Then, decay of the initial perturbation of special type corresponding to an oscillating slowly damped dipole mode for $W_0 = 5$ sphere was modelled with $N$-body simulations.

The modes referred to in the cited paper were obtained by the analytic continuation of the left side of dispersion  equation (DE) ${\cal D}(\omega)=0$.
For the perturbed functions  depending on time as $\exp (-\ii \omega t)$, the continuation is made from the upper complex frequency half-plane $\omega$ to the lower one, by deforming the integration contour so that it passes below the solution $\omega=\omega_{\rm L}$ of the DE. There are infinitely many such damped solutions, known as Landau-damped waves.

Matrix equations for spherical systems are cumbersome, which often makes it difficult to understand the physical side of the problem. Leaving aside for a while technical difficulties associated with the continuation of the matrix DE for spheres to the lower half-plane and subsequent modelling, we want to make a few remarks about the damped solutions, on the example of the homogeneous stellar medium.

The goal of this paper is to demonstrate that: (i) a damped mode may indeed exist, but it cannot be found from a DE continued to the lower half-plane; (ii) ‘eigenmodes’ corresponding to Landau-damped solutions are the not true, or genuine, modes defined on the real v-axis, however, they can be treated as modes on contours in complex $v$-plane; (iii) to study the Landau-damped waves, it is necessary to use the initial functions without singularities in the complex plane, i.e. so-called `entire' functions, or at least function with singularities located low enough (in the complex $\omega$-plane) in order not to interfere with the Landau damping.

Our analysis compares the initial value problem and eigenvalue approaches, for Maxwell background distribution function (DF). Contrary to plasma physics, where this model is stable at all scales, the stellar medium is unstable for large-scale perturbations.

The paper is organised as follows. Section 2 contains basic equations and proves the existence of the damped mode in the case when the stellar medium is unstable. Section 3 brings examples of deviations from standard Landau exponential damping. In Section 4 we give analytical arguments to support our numerical findings. Final Section 5 discusses implications and outlines our plans in this field. In the end, we give two Appendices in which some more specific issues are addressed.

\section{The concealed mode}

Instability of infinite homogeneous stellar medium can be approached by applying a small amplitude plane-wave perturbation
\be
	f_1(x, v, t) = f(v,t) \, \e^{\ii k x}
\ee
to the unperturbed background DF, $F_0(v)$. Throughout the paper, the unperturbed DF is one-dimensional Maxwell distribution,
\be
	F_0(v) = \rho_0 {\cal M}_{\sigma_0}(v)\,,\quad {\cal M}_\sigma(v) \equiv \frac{1}{\sqrt{2\pi}\sigma}\,\exp \Bigl(-\frac{v^2}{2\sigma^2}\Bigr)\,.
\ee
It is known that the perturbation is unstable to Jeans instability, if $k < k_J \equiv (4\pi G \rho_0)^{1/2}/\sigma_0$~\citep[e.g.,][hereafter BT]{BT08}. Here we use standard notation: coordinate $x$-axis is directed along wavevector $\vk$, wavenumber $k = |\vk|$, $k_J$ is a so-called Jeans wavenumber, $\rho_0$ and $\sigma_0$ are constant background density and velocity dispersion, $G$ is the gravitation constant. It is convenient to adopt units in which $4\pi G = \rho_0 = \sigma_0= 1$, so that the wavenumber are now measured in $k_J$; velocities $c$, $v$, and $u$ in $\sigma_0$; frequencies and growth/damping rates in Jeans frequency $\Omega_J \equiv (4\pi G \rho_0)^{1/2}$.

\subsection{The initial problem}

By linearising the collisionless Boltzmann and Poisson equations, it is easy to obtain the equation governing time evolution of the perturbed DF:
\be
   \frac{\p f(v,t)}{\p t} = -\ii k\Bigl[v\,f(v,t) + \eta_k(v) \rho(t) \Bigr]\,,
   \label{eq:hsm-ev}
\ee
where
\be
   \eta_k(v) \equiv \frac{4\pi G}{k^2}\, F'_0(v)\,,
   \label{eq:eta}
\ee
and $\rho(t)$ is a perturbed density,
\be
   \rho(t) \equiv \ilii \d u\, f(u,t) \,.
   \label{eq:dens}
\ee

A similar initial problem with $f(v,0) = g(v)$ in plasma was first treated by \cite{Landau_1945} analytically using inverse Laplace transform. Rewriting his eqs. (10) and (12), one can have:
\be
   \rho(t) = \frac{1}{2\pi\ii} \il_{-\infty + ic_*}^{\infty+ic_*} \d c\, \rho_c\, \e^{-\ii c k t} \,,
   \label{eq:dens_ilt}
\ee
where $c = \omega/k$ is the complex phase velocity of the wave,
\begin{align}
   &\rho_c = \frac{1}{{\cal D}^+_k(c)} \il_\curvearrowbotright\d u\, \frac{g(u)}{u-c} \,,   \label{eq:dens_lt} \\[2mm]
   &{\cal D}^+_k(c) \equiv 1 + \il_\curvearrowbotright 
   \d u\, \frac{\eta_k(u)}{u-c}\,,    \label{eq:hsm-den}
\end{align}
$c_*$ is a constant chosen so that all singularities of $\rho_c$ are located in the half-plane $\Im (c) < c_*$, symbol `$\curvearrowbotright$'
    denotes the Landau integration contour passing {\it below} the singularity at $u=c$.\footnote{In BT, this integration contour is denoted as ${\cal L}$.} Superscript `+' in (\ref{eq:hsm-den}) indicates that this function is defined in the upper half-plane and continued analytically to the lower half-plane.

\begin{figure}
\centering
  \centerline{\includegraphics[width=\linewidth, clip=]{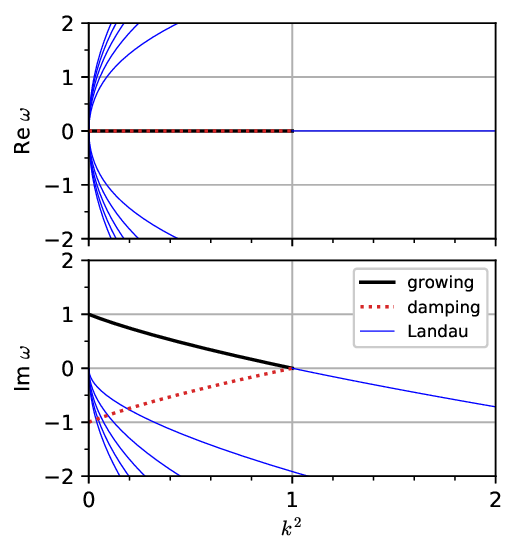} }
  \caption{Unstable (solid black) and Landau-damped (thin blue) solutions, in terms of $\omega = ck$, of the dispersion relation (\ref{eq:hsm-de}), and the damped mode (red dots), cf. BT, Fig. 5.2. The stellar medium is unstable to Jeans instability for $k<1$.}
  \label{fig:hsm}
\end{figure}

If the initial perturbation is given by an entire function $g(v)$, i.e. it has no singularities for finite complex $v$, the integral in (\ref{eq:dens_lt}) has no singularities in complex $c$\,-plane, and behaviour of $\rho(t)$ is determined by zeros of the denominator. Thus, we obtain the well-know dispersion relation (DR):
\be
   {\cal D}^+_k(c) = 0 \,.
   \label{eq:hsm-de}
\ee
Solution to this relation is given in Fig.\,\ref{fig:hsm} \citep[][BT]{1974PThPh..52.1807I}. For $k<1$, it consists of an aperiodic growing mode with growth rate $\gamma_k \equiv \Im\omega$ and many so-called Landau-damped waves describing exponential decay for density (but not for the perturbed DF, see Sect. 2.2). In the stable domain, $k>1$, the growing mode is replaced by aperiodic Landau solution. For convenience, we shall refer to the damping rate of the aperiodic Landau-damped solution, which continues the growing mode in the stable domain, as $\gamma^{(0)}_{\rm L}$, and all other (oscillating) Landau-damped solutions as $\gamma^{(j)}_{\rm L}$, $j=1, ...$, (all $\gamma_L^{(j)}<0$)  in ascending order of the damping rate.

\subsection{The eigenvalue problem}

\begin{figure}
\centering
  \centerline{\includegraphics[width=\linewidth, clip=]{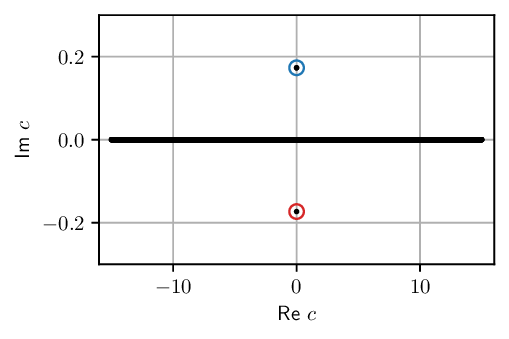} }
  \caption{Spectrum of modes in the complex phase velocity space $c$ obtained from the matrix equation for infinite homogeneous stellar medium, $k=0.9$. The unstable discrete mode of Jeans instability is marked by the blue circle. The concealed damped mode is marked by the red circle. The overlaping black dots on the real $c$-axis present a continuum spectrum of van Kampen modes.}
  \label{fig:sp}
\end{figure}

In the eigenvalue problem, we seek for solutions in the form:
\be
   f(v,t) = \tilde f(v) \e^{-\ii ck t} \,
   \label{eq:hsm-ev1}
\ee
with complex phase velocity $c$ to be determined. From (\ref{eq:hsm-ev}) and (\ref{eq:dens}) one obtains:
\be
   c \tilde f(v) = v \tilde f(v) + \eta_k(v) \ilii \d u\, \tilde f(u) \,,
   \label{eq:hsm-mat}
\ee
which can be easily solved using matrix approach \cite[see, e.g.][]{EP04, EP05, 2018MNRAS.478.4268P}. The solution for a given $k$ consists of spectrum of modes, presented in Fig.\ref{fig:sp}. In the unstable $k$-domain, there are two discrete modes: the growing mode $c_+ = \ii\gamma_k/k$ already known from Fig.\,\ref{fig:hsm}, a complex conjugate damped mode $c_- = -\ii\gamma_k/k$, and a continuum spectrum of so-called van Kampen modes~\citep{1955Phy....21..949V}. The discrete modes are absent in the stable domain $k>1$. It is widely believed that the Landau-damped waves can be regarded as a superposition of van Kampen modes (e.g., BT, p. 415). This is true only for $k>1$. For $k<1$ the presence of discrete modes makes the situation more complicated, see Sect. 3.2.

Let's turn our attention now to the damped mode marked by a red circle in Fig.\,\ref{fig:sp}. For a long time this mode was regarded as an extraneous solution, because it satisfies a relation
\be
   D_k^-(c)=0\,,
   \label{eq:hsm-drc}
\ee
rather than (\ref{eq:hsm-de}), with
\be
  {\cal D}^-_k(c) \equiv 1 + \il_\curvearrowtopright \d u\, \frac{\eta_k(u)}{u-c}\,   \label{eq:hsm-drm}
\ee
and the integration contour now passes {\it above} the singularity at $u=c$ (e.g., along the real $u$-axis for $\Im (c) <0$). There are two reasons why this solution should be treated seriously.


First, \cite{1959AnPhy...7..349C} has found explicit forms of the eigenfunctions for both discrete modes and van Kampen modes and proved that they are complete and orthogonal, in a sense that any function can be represented uniquely as superposition of their eigenfunctions:
\be
   g(v) =  b_+\, g_+(v)+b_-\, g_-(v) + g_{\rm vK}(v) \,.
   \label{eq:expansion}
\ee
In our notations, an eigenfunction corresponding to the unstable mode $c_+$ is
\be
   g_+(v) = - \frac{\eta_k(v)}{v-c_+}\,.
   \label{eq:g1}
\ee
Since $c_+$ obeys the relation (\ref{eq:hsm-de}) and $\Im (c_+) >0 $, it follows from (\ref{eq:hsm-den}) that  the eigenfunction is normalised to unity. An eigenfunction corresponding to the damped mode $c_-$ is a complex conjugate to $g_+$:
\be
   g_-(v) = g_+^*(v) = - \frac{\eta_k(v)}{v-c_-} \,,
   \label{eq:g2}
\ee
which obviously means that it shares the same normalisation, i.e.:
\be
   \ilii \d u\, g_\pm (u) = 1\,.
   \label{eq:gi_norm}
\ee
Function
\be
   g_{\rm vK}(v) \equiv \ilii \d c\, B(c)\,g_c(v)
   \label{eq:vk}
\ee
represents a superposition of van Kampen modes,
\be
  g_{c}(v) \equiv -{\cal P}\,\frac{\eta(v)}{v-c} + \lambda(c)\,\delta(v-c)\,,
\ee
where ${\cal P}$ denotes the Cauchy principal value, $\delta(v)$ is the Dirac delta function, $\lambda(c)$ is needed to satisfy normalisation of $g_{c}(v)$ to unity, i.e.:
\be
   \ilii \d u\, g_{c}(u) = 1\,,
   \label{eq:gcn}
\ee
from where
\be
   \lambda(c) = 1+{\cal P}\,\ilii \d u\,  \frac{\eta_k(u)}{u-c}\,.
   \label{eq:lam}
\ee
Given an initial profile for the perturbation $g(v)$, the expansion coefficients are obtained from the following expressions:
\begin{align}
   &b_\pm = -\frac1{C_\pm} \ilii \d u\, \frac{g(u)}{u-c_\pm}\,,\qquad C_\pm = \ilii \d u\, \frac{\eta_k(u)}{(u-c_\pm)^2}\,, \label{eq:bj} \\[2mm]
   &B(c) = \frac{1}{\lambda^2(c) + \pi^2 \eta_k^2(c)} \left[ \lambda(c)\, g(c) - \eta_k(c)\, {\cal P}\!\! \ilii \d u\, \frac{g(u)}{u-c} \right]\,.
   \label{eq:B}
\end{align}
In particular, it can be shown that $B(c)=0$ and $b_+=0$, if $g(v)=g_-(v)$.

The eigenfunction of the damped mode obtained from the matrix equation (\ref{eq:hsm-mat})  coincides with function (\ref{eq:g2}).

\begin{figure}
\centering
  \centerline{\includegraphics[width=\linewidth, clip=]{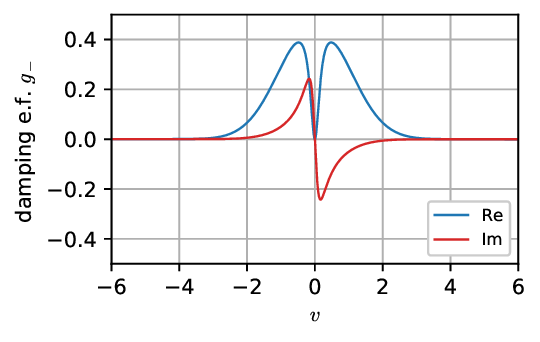} }
  \caption{Eigenfunction of the damped mode $g_-(v)$, $k=0.9$. Numerical solution of (\ref{eq:hsm-ev}) with this initial condition gives $f(v,t) = g_-(v) \exp(-\gamma_k t)$, i.e. the shape of the perturbed DF is preserved.}
  \label{fig:def}
\end{figure}

Second, it can be shown numerically that initial condition $g(v) = g_-(v)$ gives rise to $f(v,t) = g_-(v) \exp(-\gamma_k t)$. In other words, the shape of the perturbed DF is preserved (Fig.\,\ref{fig:def}), which is a characteristic of a genuine mode.
Physically, it is quite obvious that eigenfunctions of the discrete modes used as initial states for eq. (\ref{eq:hsm-ev}) lead to exponential density growth/decay with rate $\pm\gamma_k$. In Appendix A we show this explicitly using (\ref{eq:dens_ilt}) and (\ref{eq:dens_lt}).

The two reasons considered above demand to complement Fig.\,\ref{fig:hsm} by the damped mode (see red dots in both panels).

\begin{figure}
\centering
  \centerline{\includegraphics[width=\linewidth, clip=]{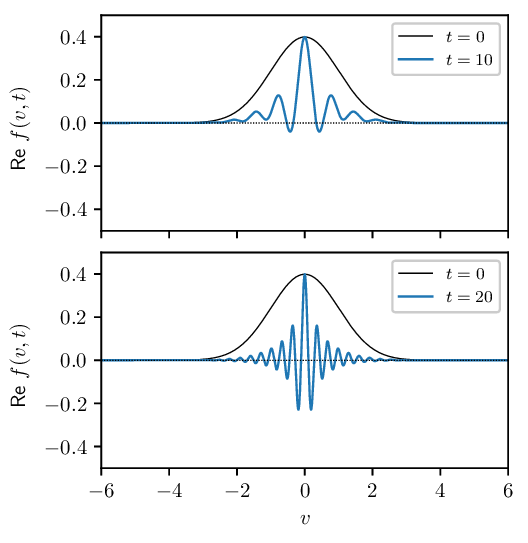} }
  \caption{Deformation of the initial Maxwell distribution (thin black lines) with time ($k=1.1$). Upper/lower panel shows real part of $f(v,t)$ at time $t=10$/$t=20$. The DF is gradually sheared out while density (\ref{eq:dens}) decreases exponentially in accordance with Laudau damping rate.}
  \label{fig:lds}
\end{figure}

Note that the exponential density damping, which is shown by both genuine modes and Landau-damped waves, manifests itself differently in the behaviour of the DFs in the two classes of solution. For the formers, shapes of the DF profiles do not change, but its amplitude varies proportionally to $\exp(\pm\gamma_k t)$. In the latter, the amplitudes do not change, but the shape becomes more and more jagged, see Fig.\,\ref{fig:lds}. To disentangle from genuine modes, we call them below `quasi-modes'.

\begin{figure}
\centering
  \centerline{\includegraphics[width=\linewidth, clip=]{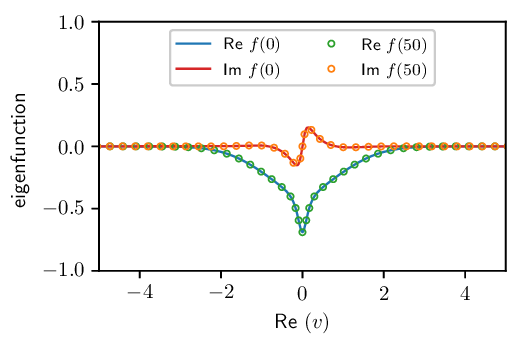} }
  \caption{Time evolution of the `eigenfunction' of the form (\ref{eq:g1}) with $c_+$ replaced by the aperiodic quasi-mode $c^{(0)} = -0.1483\ii$ ($k=1.1$), defined on the complex contour passing below $c_+$. The perturbed DF at $t=50$ is multiplied by $\exp(|c^{(0)}| k t) = 3477$ for comparison with the initial DF $f(0)$.}
  \label{fig:efld}
\end{figure}

Similarly to the initial Maxwell DF, the initial state of the form (\ref{eq:g1}) with $c_+$ replaced by the aperiodic Landau solution $c^{(0)}  \equiv -\ii |\gamma^{(0)}_{\rm L}|/k$ gives rise to DF shearing , and asymptotically to density decay $\propto\exp(\gamma^0_{\rm L} t)$. This takes place if the DF and other entries of (\ref{eq:hsm-ev}) are defined on the real $v$-axis. Now consider the task on the complex $v$-contour passing below $c^{(0)}$, e.g.:
\be
	v_{\rm I} = -2 |c^{(0)}|\,\exp\left(-\case{1}{2}\,v_{\rm R}^2 \right)\,,
\ee
where $v_{\rm R}$ and $v_{\rm I}$ are the real and imaginary parts of velocity $v$.
The corresponding DF is shown in Fig.\,\ref{fig:efld} with solid lines. Its time evolution is just decreasing of the amplitude preserving the shape of the function. To demonstrate this, we give DF at $t=50$ multiplied by $\exp(|c^{(0)}| k t)$ (`$\circ$'-marks). We conclude that constructed DF is a genuine eigenmode, {\it but on the complex contour!}  Corresponding `density' defined as intergal $\int f(v,t)\,dv$ over this complex contour decays strictly exponentially from the very beginning.

To sum up, this section argues that:
\begin{itemize}
	\item the matrix method on the real $v$-axis gives discrete complex conjugate pairs and a proxy\footnote{In any matrix approach, the continuum spectrum is represented by a finite number of matrix eigenvalues, and the singular eigenfunctions are replaced by corresponding eigenvectors.} to van Kampen modes. The corresponding eigenfunctions do not change their shapes -- a characteristic of genuine modes. Any initial perturbations can be expanded over these modes;
	\item Landau-damped waves are not true modes -- they don't have eigenfunctions on the real $v$-axis. A perturbation decays in mean, i.e. exponential decay takes place for the perturbed density, not for the perturbed DF;
	\item Landau-damped waves do have `eigenfunctions' on a complex $v$-contour passing below the corresponding zero of ${\cal D}_k^+(c)$.
\end{itemize}

\section{Deviations from Landau damping}

In this section, we give numerical evidence of deviations from expected exponential decay of Landau-damped waves. In particular, we show that superposition of van Kampen modes may lead to density decay of various types.

The solution shown in Fig.\,\ref{fig:lds} is for initial Maxwell distribution $g(v) = {\cal M}_{1}(v)$, i.e. for an entire function. The Landau damping would appear as usual if singularity of $g(v)$ was below $-\ii\gamma^0_{\rm L}/k$.
A more peculiar decays occur when the initial $g(v)$ is set using the expansion function $B(c)$ for van Kampen modes. The needed expression reads:
\be
   g(v) = B(v) + {\cal P}\! \ilii \d c\, \, \frac{B(v)\,\eta_k(c) + B(c)\,\eta_k(v)}{c-v}\,.
   \label{eq:fB}
\ee
In a sense, it is the inverse of eq. (\ref{eq:B}).

\subsection{Stable domain, $k>k_J$}

\begin{figure}
\centering
  \centerline{\includegraphics[width=\linewidth, clip=]{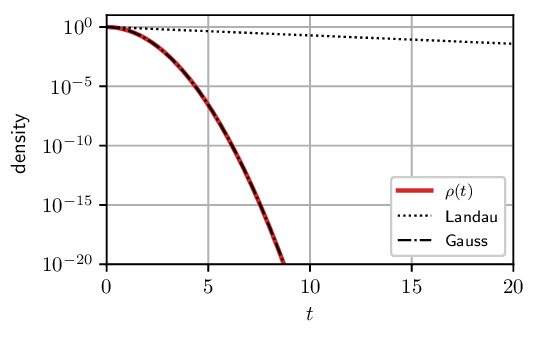} }
  \caption{Decay of the initial state given by Maxwell distribution of van Kampen waves, $B(c) = {\cal M}_{\sigma}(c)$. $f(v,t)$ given by (\ref{eq:fB}) decays as a Gaussian in time $\exp(-\sigma^2 k^2 t^2/2)$ (black dash-dotted line), not like the Landau-damped wave (black dots) for $\sigma=1$, $k=1.1$. }
  \label{fig:gt1}
\end{figure}

First of all, we apply eq. (\ref{eq:fB}) to typical profiles of choice -- Maxwell and Lorentz. Fig.\,\ref{fig:gt1} shows density decay for $B(c) = {\cal M}_1(c)$, which turns out to be perfect Gaussian in time. The Landau damping (black dots) is much slower.
This numerical result can be easily confirmed analytically. Indeed, each of the van Kampen modes evolves with its own frequency $\omega = ck$, so
\be
	 f(v,t) = \ilii \d c\, B(c)\,g_c(v) \exp (-\ii ck t)\,.
\ee
For density (\ref{eq:dens}), one obtains using normalisation of $g_c(v)$ from (\ref{eq:gcn}):
\be
	 \rho(t) = \ilii \d c\, B(c)\, \exp (-\ii ck t)\,.
\ee
Substituting Maxwell distribution $B(c) = {\cal M}_{\sigma}(c)$, one obtains the found fit. For Lorentz distribution:
\be
	B(c)=\frac{1}{\pi}\,\frac{\sigma}{c^2+\sigma^2}
\ee
one finds an exponential density decay with rate $k\sigma$, rather than Landau damping rate $\gamma^{0}_{\rm L}$.

\begin{figure}
\centering
  \centerline{\includegraphics[width=\linewidth, clip=]{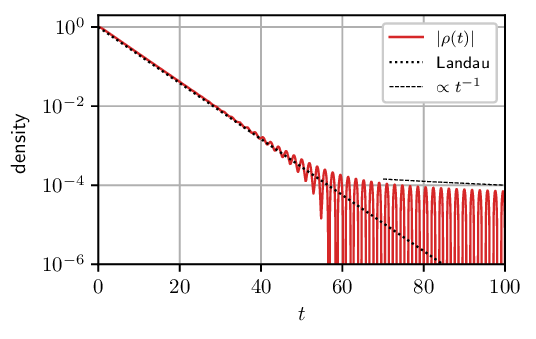} }
  \caption{Decay of Maxwell initial state $g(u)={\cal M}_{1}(u)$ with $3\sigma$ cut-off, $k=1.1$:  Landau damping (at rate $\gamma_{\rm L}^{(0)} = -0.187$, black dots) is changed by the power-law damping (dashes). }
  \label{fig:gt2}
\end{figure}

Next, Fig.\,\ref{fig:gt2} presents the density decay in the case when initial Maxwell distribution is cutted above $v_*$:
\be
	g(v) = {\cal M}_{1}(v)\qquad {\rm for}\quad |v|<v_* = 3\,,
\ee
and zero otherwise. After a short, barely visible transition period $\Delta t \sim 1$, the decay starts with Landau damping rate, but eventually power-law decay $\propto \sin(kv_*t)/t$ overtakes. Note that a similar asymptotical power-law behaviour accompanied by oscillations was found by \citet{2011JPhA...44N5502B}. These authors studied evolution of perturbations in one-dimensional non-homogeneous medium with artificial potential when action $J$ varies in semi-finite interval $J_a<J$, and obtained the density decay  $\rho \, \propto\, \exp [-\ii m \Omega(J_a) t]/t^n$, where the integer positive index $n$ ($n=1$, 2, or 3) depends on the form of the initial disturbance.

\subsection{Unstable domain, $k<k_J$}

In general, initial state (\ref{eq:expansion}) contains the exponentially growing mode leading to the density change in time:
\be
	\rho(t) =  b_+ \e^{ \gamma t} + b_- \e^{-\gamma t} + \ilii \d c\,B(c) \,\e^{-\ii kct}
	\label{eq:rho-t}
\ee
(see Appendix A). Study of the damped solutions thus requires elimination of this mode from the initial state. So, for an arbitrary $g(v)$, we consider the initial distribution:
\be
	f(v,0) = g(v) - b_+ g_+(v) = g_{\rm vK}(v) + b_-\, g_-(v)\,.
\ee
It still consists of contributions of van Kampen modes and the discrete damped mode.

One would naturally expect that if the damping rate of the discrete mode is smaller than the Landau damping rate, i.e. $|\gamma_k| < |\gamma^1_{\rm L}|$ (it holds for $k > k_* \approx 0.437$), density for this initial condition will decay $\propto \exp(-\gamma_k t)$. On the other hand, it is believed that pure superposition of van Kampen modes leads to density decay with the Landau damping rate (e.g., BT, p. 415).

\begin{figure}
\centering
  \centerline{\includegraphics[width=\linewidth, clip=]{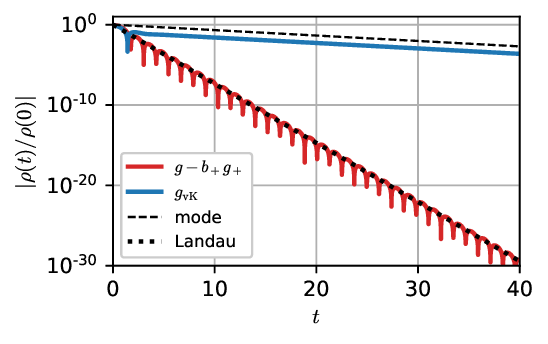} }
  \caption{Density decay of two initial states, $k=0.9$: (i) Maxwell $g(v)={\cal M}_{1}(v)$ with growing mode $b_+g_+(v)$ subtracted (solid red) decays with the Landau damping rate, $\omega_{\rm L}^{(1)} = \pm 2.581 - 1.694\ii$; (ii) a superposition of van Kampen waves $g_{\rm vK}(v)$ decays with a rate of the damped  discrete mode, $\gamma_k= -0.1558$ (solid blue).}
  \label{fig:j1}
\end{figure}

Direct evaluation of (\ref{eq:hsm-ev}) for $k=0.9$ presented in Fig.\,\ref{fig:j1}, however, demonstrate quite the opposite. The red curve marking the solution for initial DF $g-b_+g_+$ decays with Landau damping rate $\gamma^{(1)}_{\rm L}$. The oscillations of the density occur because the corresponding Landau solution has a nonzero real part of the frequency. The blue curve for initial DF $g_{\rm vK}(v)$, after some transition period, decays $\propto\exp(-\gamma_k t)$. Note that although the damping rate of the density decay coincides with the damping rate of the discrete mode, the character of this decay is the same as for quasi-modes, see Fig.\,\ref{fig:lds}.

\begin{figure}
\centering
  \centerline{\includegraphics[width=\linewidth, clip=]{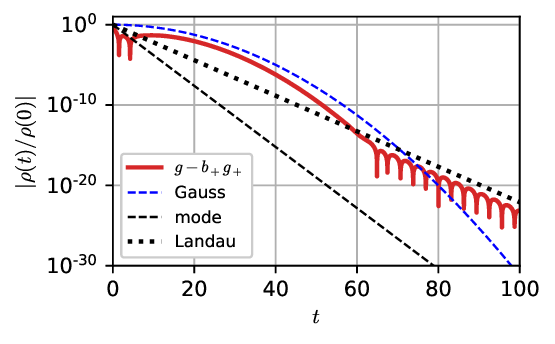} }
  \caption{Density decay of Maxwell initial state $g(v)={\cal M}_{0.4}(v)$ with growing mode $b_+g_+(v)$ subtracted (solid red) for $k=0.3$. `Gauss' blue dashed curve shows Gaussian in time decay $\exp(-\sigma^2 k^2 t^2/2)$, black dashes show dicrete mode decay $\exp(-\gamma_k t)$, black dots show Landau-damped decay $\exp(-\gamma^{(1)}_{\rm L} t)$. Here $\gamma_k = 0.8761$; $c_k = 2.92032\ii$; $\omega^{(1)}_{\rm L} = \pm 1.0083 - 0.5090\ii$, $c_{\rm L}^{(1)}=\pm 3.36087 -1.69663\ii$.}
  \label{fig:j2}
\end{figure}

In domain $k < k_* $ both initial states predictably decay with rate $\gamma^{(1)}_{\rm L}$ in agreement with Landau theory. Nevertheless, deviations could happen here as well, for example when considering narrow initial DFs. Fig.\,\ref{fig:j2} shows a long transition period for initial $g(v)={\cal M}_{0.4}(v)$ (with the growing mode subtracted) for $k=0.3$. The transition is approximately Gaussian decay $\propto \exp(-\sigma^2 k^2 t^2/2)$, which is replaced by the Landau damping at $t \sim 2|\gamma^{(1)}_{\rm L}| \sigma^{-2} k^{-2}$.
%
%

To sum up, this section shows:
\begin{itemize}
	\item construction an initial perturbation from van Kampen waves only by defining function $B(c)$ allows to obtain various decaying laws that have nothing to do with Landau damping (yet, function $g(v)$ is smooth on the real axis);
	\item cut-off of the initial function above some value in the velocity space leads to power-law decay (power law is known to appear also if $g(v)$ is not smooth);
	\item if an initial perturbation is given by an entire function, its van Kampen part does not decay with Landau damping rate in the unstable $k$-domain;
	\item a transition process before Landau damping regime could be lengthy (in our case, we observe Gaussian in time density decay because the initial $B$-distribution is Gaussian).
\end{itemize}

\section{Puzzle solving}

In the previous sections, we stumbled upon some surprising numerical evidence concerning the time evolution of perturbations in the homogeneous stellar medium.

\medskip
\noindent
{\bf 1. The damped mode.} Matrix equation (\ref{eq:hsm-mat}) predicts the existence of the damped mode, see Figs.\,\ref{fig:sp},\,\ref{fig:def}. The question arises why this solution is missed in the standard approach (e.g., BT).
The task considered by Landau was to find the evolution of an arbitrary initial perturbation given by an entire function. It evolves as the sum of contributions corresponding to the singularities of (\ref{eq:dens_lt}), or zeros of ${\cal D}_k^+(c)$. These contributions have an exponentially growing component and components corresponding to Landau-damped waves, but there is no damped mode in this expansion. One might think that the damped mode is incorporated in the Landau-damped waves.

Nevertheless, this mode exists, as confirmed by our numerical solution of evolutionary eq. (\ref{eq:hsm-ev}). It can be found from DE involving integration along the real $u$-axis, or (\ref{eq:hsm-drm}). The latter provides solutions complex conjugate to (\ref{eq:hsm-de}), and plays the same role as (\ref{eq:hsm-de}) under time reversal. The damped mode forward in time appears as the growing mode when integrating backwards in time.

\medskip
\noindent
{\bf 2. Density decay in the unstable $k$-domain}. It is naturally expected that a packet of van Kampen modes
\be
   g_{\rm vK}(v) = \ilii \d c\, B(c)\, g_c(v) = g(v) - b_+ g_+(v) - b_- g_-(v)
\ee
decays with Landau damping rate. However this is not the case when $|\gamma_k| < |\gamma^{(1)}_{\rm L}|$, as is evident in Fig.\,\ref{fig:j1}. We shall show that this is due to a singularity of $B(c)$ at $c=c_-$.

Since each of the van Kampen modes, $g_c$, evolves in time $\propto\exp(-\ii kct)$, their contribution to the total density is:
\be
	\rho_{\rm vK}(t)= \ilii \d c\, B(c)\,\exp(-\ii kct)\,.
	\label{eq:vkm}
\ee
For positive $kt$, the integration can be performed over a contour closed in the lower half-plane and replaced by a sum of residues:
\be
	\rho_{\rm vK}(t)= -2\pi\ii \sum\limits_n {\rm Res} (c_n)\,\exp(-\ii k c_n t)\,.
\ee
Here $c_n$ are all poles of $B(c)$ in the lower half-plane. In (\ref{eq:B}) this function is defined on the real $c\,$-axis. Our goal is to prove that an analytic continuation of $B(c)$ to the lower half-plane $c$, apart from poles of Landau quasi-modes, contains also a pole due to the damped mode $c = c_-$, and near this pole $B(c)$ has the form:
\be
    B(c)\approx \frac{1}{2\pi\ii} \frac{b_-}{c-c_-}\,,
	\label{eq:hsm-x1}
\ee
where
\be
    b_-=- \ilii\dfrac {g(u)\,\d u}{u-c_-} \left[ \ilii \dfrac{\eta_k(u)\,\d u}{(u-c_-)^2} \right]^{-1} .
	\label{eq:hsm-x2}
\ee

For real $c$, expression (\ref{eq:B}) can be decomposed, with the aid of Eq.\,(\ref{eq:lam}),  as follows:
\be
   B(c)= B^+(c) + B^-(c)\,,
   \label{eq:exp-B}
\ee
where
\begin{align}
   &B^\pm(c) \equiv \frac{g^\pm(c)}{\varepsilon^\pm(c)}\,,\\[2mm]
   &g^\pm(c)=\frac{g(c)}{2}\pm\frac1{2\pi \ii}{\cal P}\ilii \frac{g(u)\,\d u}{u-c}\,,\\[2mm]
   &\varepsilon^\pm(c)=1+2\pi\ii \left[\pm\frac{\eta_k(c)}{2}+\frac1{2\pi\ii} {\cal P}\ilii \frac{\eta_k(u)\,\d u}{u-c}\right].
\end{align}
Now analytic continuation is obvious, since $\varepsilon^\pm(c)$ can be replaced by ${\cal D}_k^\pm(c)$ off the real axis, and $g^\pm$ are replaced by
\be
g^+(c) =  \frac1{2\pi\ii} \il_\curvearrowbotright \frac{g(u) \d u}{u-c}\,,\ \ \ \ \
g^-(c) =-\frac1{2\pi\ii} \il_\curvearrowtopright \frac{g(u) \d u}{u-c}\,,\ \ \ \ \
\ee
where, depending on signs `$\curvearrowbotright$' or `$\curvearrowtopright$', integration contour passes below or above the singularity at $u=c$. Note that
\begin{multline}
   g^+(c)+ g^-(c) = \frac1{2\pi\ii} \Bigl[ \il_\curvearrowbotright  \frac{g(u)\,
   \d u}{u-c} -\! \il_\curvearrowtopright  \frac{g(u)\, \d u}{u-c}\!\Bigr]\\
    = \frac1{2\pi\ii} \oint \frac{g(u)\, \d u}{u-c}=g(c)\,,
\end{multline}
i.e. this is a decomposition of function $g(v)$.
Since ${\cal D}_k^-(c)$ has one zero in the lower half-plane $c=c_-$, $B^-(c)$ has a pole at this point leading to $\propto \exp(-\gamma_k t)$ contribution to density decay. Expansion ${\cal D}_k^-(c)$ near $c=c_-$ gives:
\be
   {\cal D}_k^-(c) \approx \left.\frac{\d {\cal D}_k^-(c)}{\d c}\right|_{c_-} (c-c_-) =  (c-c_-) \il_\curvearrowtopright \d u \frac{\eta_k(u)}{(u-c_-)^2}\,.
\ee
Finally, substitution to
\be
    B^-(c) = \frac{g^-(c)}{{\cal D}_k^-(c)}
\ee	
leads to the desired result (\ref{eq:hsm-x1}), if we choose the integration contour along the real $u$-axis. The input from this pole gives a density decay slower that the Landau damping for $k>k_*\approx 0.437$, as is seen in Fig.\,\ref{fig:j1}.

Note that decomposition (\ref{eq:exp-B}) is not unique. Moreover, it is possible to decompose $B(c)$ so that parts of new decomposition have no singularities in the upper/lower half-planes (Appendix~B).

Gaussian in time density decay occurred for Maxwell $B(c)$ (Fig.\,\ref{fig:gt1}) is simply a reflection of the fact that $B(c)$ has no singularities in the lower half-plane.

\begin{figure}
\centering
  \centerline{\includegraphics[width=\linewidth, clip=]{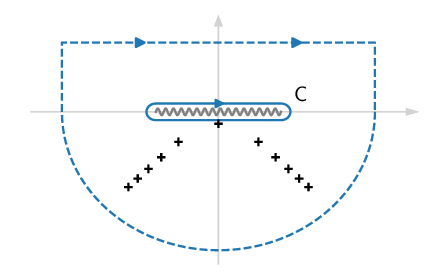} }
  \caption{Integration contour in (\ref{eq:dens_ilt}) in the complex $c\,$-plane (blue dashes).  '+' signs mark the Landau-damped quasi-modes obtained as zeros of ${\cal D}_k^+(c)$. A wavy line marks the cut on real axis $c$:  $-v_*<c<v_*$.}
  \label{fig:lc}
\end{figure}

\medskip
\noindent
{\bf 3. Power-law decay.} To understand an appearance of the power-law term (Fig.\,\ref{fig:gt2}), we should reexamine derivation of eqs. (\ref{eq:dens_lt}, \ref{eq:hsm-den}). They are obtained as a result of analytic continuation of the integrals over the real $u$-axis by deformation of the integration contour to the lower half-plane. Since $g(v)$ is zero for $|v|>v_*$, the cut in the complex plane is finite (Fig.\,\ref{fig:lc}), and deformation of the contour is not needed for this integral. On the either side of the cut
\be
	g_*(c) = \ilii \d u\, \frac{g(u)}{u- (c \pm \ii 0)} = \pm \ii \pi g(c) + {\cal P} \il_{-v_*}^{v_*} \d u\, \frac{g(u)}{u-c} \,.
	\label{eq:g-fin}
\ee
The cut leads to an additional input to density (\ref{eq:dens_ilt}):
\be
	\rho_{\rm cut}(t) = \oint\limits_{C} \d c\, \frac{g_*(c)}{{\cal D}_k^+(c)}\, \e^{-\ii ckt}= 2\pi\ii \il_{-v_*}^{v_*} \d c\, \frac{g(c)}{{\cal D}_k^+(c)}\,\e^{-\ii ckt}\,.
\ee
Integrating by parts, assuming $g(v)$ is even and real, one obtains for large $t$:
\be
	\rho_{\rm cut}(t) = -\frac{4\pi}{kt}\, g(v_*) \, \Im \Bigl[ \frac{e^{-\ii v_* kt}}{{\cal D}_k^+(v_*)} \Bigr] + {\cal O}(t^{-2})\,.
\ee
Expression in the square brackets gives sinusoidal oscillations with some phase shift due to the presence of ${\cal D}_k^+(v_*)$ in the denominator. If $g(v_*)$ is small, the power-law decay reveals itself only after a while.


\section{Implications}
\label{sec:summary}

In this article, we analysed numerical solutions of equations describing perturbations of the homogeneous stellar medium. Although the equilibrium state is problematic in the infinite media and invokes a so-called Jeans swindle (e.g., BT), it is a good testing ground for many phenomena in stellar dynamics~\citep[see][for example of a more realistic setup]{1971SvA....14..758B}. The unexpected behaviour of damped solutions in the unstable domain is associated with the presence of a genuine damped mode, which is complex conjugate to the unstable Jeans mode. The damped mode cannot be found as a solution to the analytic continuation of DR from upper complex frequency half-plane $\omega$ to the lower half-plane. It is a solution to another DR, in which the integration is carried out over real phase variables (velocity $u$ in the case of the homogeneous medium and action variables in the case of spheres). This mode is usually ignored (cf. e.g., BT, Fig.\,5.2 and our Fig.\,\ref{fig:hsm}).

We have shown that Landau-damped waves possess their `eigenfunctions', but defined on complex contours passing below the corresponding solution of continued DE. No eigenfunctions of Landau solutions exist on the real axis. Therefore, we call these solutions not `genuine modes', but `quasi-modes', emphasizing that exponential decay takes place only on average, i.e. for density and potential. In genuine damped modes, DF is decreasing exponentially (along with density and potential).

\citet{1994ApJ...421..481W} numerically investigated the evolution of the  initial perturbation of DF taken as `eigenfunction' corresponding to Landau-damped quasi-mode. It contains  denominator similar in shape to our solution for damped mode $g_-(v)$ (\ref{eq:g2}), but instead of velocity, real action variables appear in it. We have argued that Landau-damped solutions have no eigenfunctions as functions of real variables. 
Besides, we have seen that the use for initial DF of non-analytic functions with the pole-type singularity in the complex plane (in the velocity space or the action space in the case of spherical systems) leads to artefacts –- the appearance of the exponential decay with given characteristics. E.g., the exponential decay could be slower, as in the case of the Lorenz initial DF with sufficiently small $\sigma$ (sect. 3.1). Moreover, a simple pole $v=c_{\rm L}$ of the initial DF along with a simple zero of the function ${\cal D}_k^+(c)$ gives rise a second-order pole at $c=c_{\rm L}$ leading to $\propto \, t \exp(-\ii c_{\rm L} k t)$ asymptotical term instead of the expected exponential Landau damping.
We consider the initial conditions given by entire functions or the use methods that are not related to the choice of initial conditions with singularities to be more appropriate. An example of such a study is provided by \citet{2020MNRAS.492.6019H}.

In connection with the aforesaid, only solutions found without analytic continuation to the lower half-plane turn out to be genuine modes, e.g., for real values of the velocity $u$ in the integrals of Laplace transform (\ref{eq:dens_lt}).

In the future, we plan to extend our research from the homogeneous medium to spherical systems  using our matrix approach for spheres \citep{2007MNRAS.379..573P} and investigate the discreteness effects that inevitable in any N-body models~\citep[e.g.,][]{2020MNRAS.492.4819P}.

\section*{Acknowledgments}

The authors thank D. Heggie for his kindly report and suggestions on improving the original version of the paper.
The reported study was funded by Foundation for the advancement of theoretical physics and mathematics ``Basis'', grant \#20-1-2-33; by RFBR and DFG according to the research project 20-52-12009; by the Volkswagen Foundation under the Trilateral Partnerships grant No. 97778. The work was partially financially supported by the Ministry of Science and Higher Education of the Russian Federation (Ilia Shukhman). High precision calculations are achieved with the use of Multiprecision Computing Toolbox developed by Advanpix.

\section*{Data Availability}

Data underlying this article will be shared on reasonable request to the authors via epolyach@inasan.ru.


\begin{thebibliography}{}
\makeatletter
\relax
\def\mn@urlcharsother{\let\do\@makeother \do\$\do\&\do\#\do\^\do\_\do\%\do\~}
\def\mn@doi{\begingroup\mn@urlcharsother \@ifnextchar [ {\mn@doi@}
  {\mn@doi@[]}}
\def\mn@doi@[#1]#2{\def\@tempa{#1}\ifx\@tempa\@empty \href
  {http://dx.doi.org/#2} {doi:#2}\else \href {http://dx.doi.org/#2} {#1}\fi
  \endgroup}
\def\mn@eprint#1#2{\mn@eprint@#1:#2::\@nil}
\def\mn@eprint@arXiv#1{\href {http://arxiv.org/abs/#1} {{\tt arXiv:#1}}}
\def\mn@eprint@dblp#1{\href {http://dblp.uni-trier.de/rec/bibtex/#1.xml}
  {dblp:#1}}
\def\mn@eprint@#1:#2:#3:#4\@nil{\def\@tempa {#1}\def\@tempb {#2}\def\@tempc
  {#3}\ifx \@tempc \@empty \let \@tempc \@tempb \let \@tempb \@tempa \fi \ifx
  \@tempb \@empty \def\@tempb {arXiv}\fi \@ifundefined
  {mn@eprint@\@tempb}{\@tempb:\@tempc}{\expandafter \expandafter \csname
  mn@eprint@\@tempb\endcsname \expandafter{\@tempc}}}

\bibitem[\protect\citeauthoryear{{Antonov}}{{Antonov}}{1960}]{1960AZh....37..918A}
{Antonov} V.~A.,  1960, \azh, \href
  {https://ui.adsabs.harvard.edu/abs/1960AZh....37..918A} {37, 918 (in Russian).} Also SvA, 4, 859 (in English)

\bibitem[\protect\citeauthoryear{{Antonov}}{{Antonov}}{1962}]{1962spss.book.....A}
{Antonov} V.~A.,  1962, Vestnik Leningrad Univ., 19, 96 (in Russian). Also de Zeeuw (1987), 531 (in English)

\bibitem[\protect\citeauthoryear{{Barr{\'e}}, {Olivetti}  \&
  {Yamaguchi}}{{Barr{\'e}} et~al.}{2011}]{2011JPhA...44N5502B}
{Barr{\'e}} J.,  {Olivetti} A.,   {Yamaguchi} Y.~Y.,  2011, \mn@doi [Journal of
  Physics A Mathematical General] {10.1088/1751-8113/44/40/405502}, \href
  {https://ui.adsabs.harvard.edu/abs/2011JPhA...44N5502B} {44, 405502}

\bibitem[\protect\citeauthoryear{{Binney} \& {Tremaine}}{{Binney} \&
  {Tremaine}}{2008}]{BT08}
{Binney} J.,  {Tremaine} S.,  2008, {Galactic Dynamics: Second Edition}.
Princeton University Press

\bibitem[\protect\citeauthoryear{{Bisnovatyi-Kogan} \&
  {Zel'dovich}}{{Bisnovatyi-Kogan} \& {Zel'dovich}}{1971}]{1971SvA....14..758B}
{Bisnovatyi-Kogan} G.~S.,  {Zel'dovich} Y.~B.,  1971, \sovast, \href
  {https://ui.adsabs.harvard.edu/abs/1971SvA....14..758B} {14, 758}

\bibitem[\protect\citeauthoryear{{Case}}{{Case}}{1959}]{1959AnPhy...7..349C}
{Case} K.~M.,  1959, \mn@doi [Annals of Physics]
  {10.1016/0003-4916(59)90029-6}, \href
  {https://ui.adsabs.harvard.edu/abs/1959AnPhy...7..349C} {7, 349}

\bibitem[\protect\citeauthoryear{{Doremus}, {Feix}  \& {Baumann}}{{Doremus}
  et~al.}{1971}]{1971PhRvL..26..725D}
{Doremus} J.~P.,  {Feix} M.~R.,   {Baumann} G.,  1971, \mn@doi [\prl]
  {10.1103/PhysRevLett.26.725}, \href
  {https://ui.adsabs.harvard.edu/abs/1971PhRvL..26..725D} {26, 725}

\bibitem[\protect\citeauthoryear{{Fridman} \& {Polyachenko}}{{Fridman} \&
  {Polyachenko}}{1984}]{1984sv...bookQ....F}
{Fridman} A.~M.,  {Polyachenko} V.~L.,  1984, {Physics of gravitating systems.
  I - Equilibrium and stability}.
Springer, New York

\bibitem[\protect\citeauthoryear{{Heggie}, {Breen}  \& {Varri}}{{Heggie}
  et~al.}{2020}]{2020MNRAS.492.6019H}
{Heggie} D.~C.,  {Breen} P.~G.,   {Varri} A.~L.,  2020, \mn@doi [\mnras]
  {10.1093/mnras/staa272}, \href
  {https://ui.adsabs.harvard.edu/abs/2020MNRAS.492.6019H} {492, 6019}

\bibitem[\protect\citeauthoryear{{Ikeuchi}, {Nakamura}  \&
  {Takahara}}{{Ikeuchi} et~al.}{1974}]{1974PThPh..52.1807I}
{Ikeuchi} S.,  {Nakamura} T.,   {Takahara} F.,  1974, \mn@doi [Progress of
  Theoretical Physics] {10.1143/PTP.52.1807}, \href
  {https://ui.adsabs.harvard.edu/abs/1974PThPh..52.1807I} {52, 1807}

\bibitem[\protect\citeauthoryear{{Landau}}{{Landau}}{1946}]{Landau_1945}
{Landau} L.,  1946, J. Phys. USSR, 10, 25

\bibitem[\protect\citeauthoryear{{Polyachenko}}{{Polyachenko}}{2004}]{EP04}
{Polyachenko} E.~V.,  2004, \mn@doi [\mnras]
  {10.1111/j.1365-2966.2004.07390.x}, \href
  {http://adsabs.harvard.edu/abs/2004MNRAS.348..345P} {348, 345}

\bibitem[\protect\citeauthoryear{{Polyachenko}}{{Polyachenko}}{2005}]{EP05}
{Polyachenko} E.~V.,  2005, \mn@doi [\mnras]
  {10.1111/j.1365-2966.2005.08660.x}, \href
  {http://adsabs.harvard.edu/abs/2005MNRAS.357..559P} {357, 559}

\bibitem[\protect\citeauthoryear{{Polyachenko}}{{Polyachenko}}{2018}]{2018MNRAS.478.4268P}
{Polyachenko} E.~V.,  2018, \mn@doi [\mnras] {10.1093/mnras/sty1402}, \href
  {https://ui.adsabs.harvard.edu/abs/2018MNRAS.478.4268P} {478, 4268}

\bibitem[\protect\citeauthoryear{{Polyachenko}, {Polyachenko}  \&
  {Shukhman}}{{Polyachenko} et~al.}{2007}]{2007MNRAS.379..573P}
{Polyachenko} E.~V.,  {Polyachenko} V.~L.,   {Shukhman} I.~G.,  2007, \mn@doi
  [\mnras] {10.1111/j.1365-2966.2007.11821.x}, \href
  {http://adsabs.harvard.edu/abs/2007MNRAS.379..573P} {379, 573}

\bibitem[\protect\citeauthoryear{{Polyachenko}, {Berczik}, {Just}  \&
  {Shukhman}}{{Polyachenko} et~al.}{2020}]{2020MNRAS.492.4819P}
{Polyachenko} E.~V.,  {Berczik} P.,  {Just} A.,   {Shukhman} I.~G.,  2020,
  \mn@doi [\mnras] {10.1093/mnras/staa141}, \href
  {https://ui.adsabs.harvard.edu/abs/2020MNRAS.492.4819P} {492, 4819}

\bibitem[\protect\citeauthoryear{{Van Kampen}}{{Van
  Kampen}}{1955}]{1955Phy....21..949V}
{Van Kampen} N.~G.,  1955, \mn@doi [Physica] {10.1016/S0031-8914(55)93068-8},
  \href {https://ui.adsabs.harvard.edu/abs/1955Phy....21..949V} {21, 949}

\bibitem[\protect\citeauthoryear{{Weinberg}}{{Weinberg}}{1994}]{1994ApJ...421..481W}
{Weinberg} M.~D.,  1994, \mn@doi [\apj] {10.1086/173665}, \href
  {https://ui.adsabs.harvard.edu/abs/1994ApJ...421..481W} {421, 481}

\makeatother
\end{thebibliography}

\section*{Appendix A: Density evolution for inital DF \protect\lowercase{$g(v) = g_{\pm}(v)$}}

\renewcommand{\theequation}{A\arabic{equation}}

The eigenfunction of the growing solution is
\be
   g(v) = g_+(v) \equiv -\frac{\eta_k(v)}{v-c_+}\,,
\ee
where $c_+= \ii \gamma_k/k$ is a solution of the dispersion relation ${\cal D}^+_k(c)=0$, see (\ref{eq:hsm-den}). For the integral in (\ref{eq:dens_lt}) one obtains:
\begin{align}
 & \il_\curvearrowbotright  \frac{\d u\, g(u)}{u-c} = - \frac{1}{c-c_+} \il_\curvearrowbotright\d u \, \eta_k(u) \Bigl[ \frac1{u-c} - \frac1{u-c_+} \Bigr]  \notag \\
 &= -\frac{1}{c-c_+}\Bigl[1+\il_\curvearrowbotright \d u \,  \frac{ \eta_k(u)}{u-c}\Bigr] = -\frac{1}{c-c_+} {\cal D}^+_k(c)\,, \label{eq:gmr}
\end{align}
valid in the whole complex $c\,$-plane. In expression for $\rho_c$, functions ${\cal D}^+_k(c)$ in the numerator and denominator cancel, so one finally obtains from
(\ref{eq:dens_ilt}):
\be
   \rho(t) = - \frac{1}{2\pi\ii} \il_{-\infty + ic_*}^{\infty+ic_*} \frac{\d c}{c-c_+} \, \e^{-\ii c k t}  = \e^{\gamma_k t} \,.
\ee

\medskip
Now we consider the eigenfunction of the damped mode
\be
   g(v) = g_-(v) \equiv -\frac{\eta_k(v)}{v-c_-}\,,
\ee
where $c_-= -\ii \gamma_k/k$ is a solution of the equation
\be
   1 + \ilii \d u \frac{\eta_k(u)}{u-c} = 0\,,    \label{eq:hsm-frown}
\ee
i.e. integration is performed above the singularity $u=c_-$. An expression similar to (\ref{eq:gmr}) in the upper half-plane is
\begin{align}
 & \ilii  \frac{g(u)\d u}{u-c} = - \frac{1}{c-c_-} \ilii \d u \, \eta_k(u) \Bigl[ \frac1{u-c} - \frac1{u-c_-} \Bigr]  \notag \\
 &= -\frac{1}{c-c_-}\Bigl[1 + \ilii \d u \,  \frac{ \eta_k(u)}{u-c}\Bigr]\,,\quad \Im (c) >0\,.
\end{align}
Analytic continuation to the lower half-plane is achieved by changing the last integral over real $u$-axis to an integral over Landau contour:
\be
	\il_\curvearrowbotright \frac{g(u)\d u}{u-c} = -\frac{1}{c-c_-}\Bigl[1 + \il_\curvearrowbotright \d u \,  \frac{ \eta_k(u)}{u-c}\Bigr] =  -\frac{1}{c-c_-}{\cal D}^+_k(c)\,,
\ee
valid in the whole complex $c\,$-plane. Finally, for density we obtain:
\be
   \rho(t) = - \frac{1}{2\pi\ii} \il_{-\infty + ic_*}^{\infty+ic_*} \frac{\d c}{c-c_-} \, \e^{-\ii c k t}  = \e^{-\gamma_k t} \,.
\ee

\section*{Appendix B: $B$-decomposition analytic in upper/lower half plane}

\renewcommand{\theequation}{B\arabic{equation}}

For entire initial DF $g(v)$, we showed that $B^+(c)$ is not analytic neither in the upper half-plane (discrete growing mode), nor in the lower half-plane (Landau-damped waves). Similarly, $B^-(c)$ contains singularity in the lower half-plane (discrete damped mode) and Landau-damped waves in the upper half-plane. It is possible however to redefine expansion (\ref{eq:exp-B}) so that new parts $\mfB^\pm(c)$ will be analytic in the upper/lower half-plane, correspondingly. In order to do this, we define
\be
   \mfg^\pm(c) \equiv g^\pm(c) \pm \frac{{\cal D}_k^\pm(c)} {2\pi \ii} \left[\frac{b_+}{c-c_+} + \frac{b_-}{c-c_-}\right]
   \label{eq:mfg}
\ee
and
\be
   \mfB^\pm(c) \equiv \frac{\mfg^\pm(c)}{{\cal D}_k^\pm(c)}\,.
   \label{eq:mfB}
\ee
It is easy to verify that
\be
   \mfB(c)= \mfB^+(c) + \mfB^-(c) = B(c)\,.
\ee

Substitution of (\ref{eq:mfg}) into (\ref{eq:mfB}) gives:
\begin{align}
   &\mfB^+(c)=\frac{1}{{\cal D}_k^+(c)} \left[g^+(c)+\frac{b_+}{2\pi \ii}\,\frac{{\cal D}_k^+(c)}{c-c_+}\right]+\frac1{2\pi \ii}\,\frac{b_-}{c-c_-} , \label{eq:mfBp}\\[2mm]
   &\mfB^-(c)=\frac{1}{{\cal D}_k^-(c)} \left[g^-(c)-\frac{b_-}{2\pi \ii}\,\frac{{\cal D}_k^-(c)}{c-c_-}\right]-\frac1{2\pi \ii}\,\frac{b_+}{c-c_+}. \label{eq:mfBm}
\end{align}
For $c$ near $c_+$:
\be
	\mfB^+(c) \approx \frac{1}{{\cal D}_k^+(c)} \left[g^+(c)+\frac{b_+}{2\pi \ii}\, \left.\frac{\d {\cal D}_k^+(c)}{\d c}\right|_{c_+}  \right]\,.
\ee
From a relation analogous to (\ref{eq:hsm-x2}),
\be
	g^+(c_+) = \frac{1}{2\pi \ii} \ilii \frac{g(u)\,\d u}{u-c_+} = -\frac{b_+}{2\pi\ii} \left.\frac{\d {\cal D}_k^+(c)}{\d c}\right|_{c_+}\,,
\ee
thus $\mfB^+(c)$ is analytic in the upper half-plane. Similarly, $\mfB^-(c)$ is analytic in the lower half-plane.

To evaluate the integral (\ref{eq:vkm}), we wish to extend the integration path to a closed contour in the lower half-plane. Since $\mfB^-(c)$ has no singularities there, we obtain using (\ref{eq:mfBp}):
\begin{multline}
	\rho_{\rm vK}(t)=  \ilii \d c\, [\mfB^+(c)+\mfB^-(c)] \, \e^{-\ii kct}  \\
	= \ilii \d c\, \mfB^+(c) \, \e^{-\ii kct}
	= \ilii \d c\, \frac{g^+(c)}{{\cal D}_k^+(c)}\, \e^{-\ii kct} - b_- \e^{-\gamma_k t}\,.
\end{multline}
This form represents explicitly contributions of the Landau damping and the damped mode. The Landau term differs from (\ref{eq:dens_ilt}) in integration contour only: here the integral is taken over the real $c\,$-axis, while there the integration contour lies above singularity $u=c_+$.

\end{document}